\input epsf
\epsfverbosetrue

\documentstyle[11pt,aaspp]{article}

\righthead{Neutrino Lasing}

\input defs.tex

\begin{document}

\title{Mixed Dark Matter from Neutrino Lasing}
\author{G.D. Starkman, N. Kaiser, R.A. Malaney\\
Canadian Institute for Theoretical Astrophysics, \\
University of Toronto,\\
Toronto, ON, Canada M5S 1A7.}
% \maketitle
\begin{abstract}

We investigate the production of dark matter with non-standard
momentum distributions arising from the decay of
a relativistic massive neutrino into a lighter fermion and boson
H$\rightarrow$F$+$B. We develop the Boltzmann equations for this process,
paying particular
attention to
spin polarisation effects.
Such decays can, under certain circumstances, lead to
a process of runaway stimulated decays or `neutrino-lasing'
resulting, for suitable boson mass, in
the production of mixed-dark-matter.
We discuss a number of variations on the model; we
show how the yield of cold particles can be increased
by a simple modification to our original proposal.
We discuss the singlet-majoron model as a possible framework for
these decays.
\end{abstract}

\keywords{dark matter -- large scale structure of universe -- elementary
particles}

\section{INTRODUCTION}

In the standard  cold dark-matter (CDM) model,
 present day structure can be traced back
to primordial density fluctuations arising during inflation.
The predictions of this model agree at least
qualitatively with what is
seen in redshift surveys, bulk motions and by COBE.  However,
detailed statistical analysis of clustering on large-scales
reveals a problem: the spectrum has the wrong shape for
wavelengths $\lambda \sim 30-100$ Mpc/$h$ (Maddox {\it et al.} 1990;
Fisher {\it et al.}; 1993, Vogeley {\it et al.} 1992; Feldman {\it et al.}
1993; Ostriker 1993).

One possible modification is to arrange for a non-scale invariant
primordial spectrum --- the so called `tilted' model --- but the
general agreement between the amplitude of $\Delta T/ T$ seen by COBE
on scales of order 1000 Mpc (Smoot {\it et al.} 1992; Wright {\it et al.} 1992)
and the amplitude of large-scale structure on scales of tens of Mpc
highly constrain such tinkering with the very early universe aspects
of the model.  A more promising approach may be to explore
modifications to the matter content, and
various authors
(Shafi and Stecker, 1984; Schaefer, Shafi and Stecker, 1989;
Schaefer and Shafi, 1992; Davis, Summers and Schlegel 1992;
Taylor and Rowan-Robinson 1992; Klypin {\it et al.} 1993) have
 suggested that  a hybrid
mixed dark-matter (MDM) universe could  more
readily account for the  large-scale power observed
in galactic surveys  and recent COBE measurements of the microwave background
fluctuations.

A  significant suggestion in this regard is that of Madsen (1992)
who has proposed that
generic neutrino decay processes in the
early universe may lead to the simultaneous production of hot and cold dark
matter.
He considered the decay processes
\begin{equation}
\label{eq:reactions}
{\rm H}\rightarrow {\rm F} + {\rm  B} \ ;\ \ \bar {\rm H}\rightarrow \bar{\rm
F}
 + \bar{\rm B}
\end{equation}
where H is the heavy neutrino;
F is some very light fermion which may  or may not
be one of the known neutrinos,
and B is a  light boson with mass $\sim 100$ eV which will constitute the
dark matter.
Decaying neutrinos have been widely considered in cosmology
(eg. see Kolb and Turner 1990 and references therein),
but mostly in the context of radiative  and/or
non-relativistic decays (though see Kawasaki, Steigman and Kang {\it et al.}
1993).
Madsen (1992) considers the case where
the decay is to a weakly interacting  low-mass boson,
and  the massive neutrinos are still highly relativistic
when the free decay time equals the age of the universe.
In the decay of a non-relativistic particle the
momenta of the decay products are $\pf,\ \pb \sim \mh / 2$ and
are therefore much greater than the ambient temperature $T$.  Consequently,
the final occupation numbers of the decay products
will be $f \sim (T / \mh)^3 \ll 1$
and the decay can proceed essentially freely.
Here things are quite different; the typical momenta of the decay products
are $p \sim T$ so if the heavy particles were to start off with a
thermal distribution and decay freely into thermally decoupled species,
the occupation numbers
for the decay products would rapidly become appreciable, $f\sim 1$,
so one must allow for the inverse decay reaction.
At $T_d$
(the temperature when the time dilated free decay time
is of order the age of universe), Madsen argues that
the decays and inverse decays will bring
the H, F and B into thermal equilibrium; he
finds that over a large range of initial conditions the temperature
is below the critical temperature for the number density of bosons,
and a substantial fraction of the bosons are in a Bose condensate.

Clearly, the condensation of some fraction of the light bosons
into  zero momentum modes
would have far-reaching consequences.
The fraction of the light bosons
in the Bose condensate  could represent the CDM component of a MDM model,
with the uncondensed bosons (and also perhaps the fermions) representing
the HDM component. However,
it is not obvious that the assumption of instantaneous
thermalisation is valid.

In order to explore this issue further it is necessary to formulate
and solve the Boltzmann equations governing this process.
As we have reported
elsewhere (Kaiser, Malaney, and Starkman 1993, hereafter KMS),
one finds that under certain circumstances a
bimodal boson momentum distribution can indeed occur,
with a substantial fraction of the bosons having
very low momenta ( a \lq\lq quasi-condensate\rq\rq),
but that this occurs by a process
of `neutrino-lasing' rather than by thermalisation and Bose-condensation.
Thus, relativistic decays may form mixed dark matter
but in a particularly intriguing and unusual manner.

KMS identified two critical requirements for lasing to
occur. First, that the decays should take place while the
heavy particles are relativistic, this being necessary in order
that the decays be able to gain access to low-momentum boson states.
Second, that there should initially be a `population-inversion' with
an excess of heavy over light fermions.  The former requirement
places a lower bound on the coupling constant governing the
decay process.
The latter requirement is rather unusual and requires that
some out-of-equilibrium process takes place to `pump' the laser.
The scheme considered in KMS was to assume that both types of
light particles had
decoupled prior to quark-confinement (at a temperature
of a few hundred MeV or so).
As the previously free quarks combined into hadrons
(a process which took place at approximately constant entropy),
the reduction in the total number of degrees of freedom
increased the entropy per remaining (coupled) degree of freedom,
and therefore the abundance of the H-neutrinos relative to the
decoupled decay products.

The discussion in KMS was incomplete in several respects:
first, no derivation of the Boltzmann equations was given,
and the equations presented and solved were for the specific case
that the decays were isotropic in the rest frame of the heavy
particle.
In fact, spin polarisation of the decaying particles can have
an important impact on the lasing process, and
such effects should be explicitly included in the Boltzmann equations.
Second, no serious attempt
was made to identify the particles in the required decay scheme.

In this paper we will attempt to fill in these important gaps in the KMS
discussion. The structure of this paper is as follows:
In \S\ref{sec:boltzmann}, we give a rigorous derivation of the Boltzmann
equations,
with particular attention paid to the anisotropy of
the decays arising from the spin polarisation of the heavy neutrinos.
In \S\ref{sec:lasing} we develop the mathematical description of
neutrino lasing. We identify the regions of parameter space
in which lasing can occur, and we show how the final abundance of
cold and hot particles can be calculated analytically.
In \S\ref{sec:majoron} we place these ideas
in the context of a particular
framework -- the singlet-majoron model.
% Here we outline the circumstances
% in which spin effects can be neglected, as well as identifying a possible
%%decay scheme
% which can be embedded in this model.
In \S\ref{sec:variations} we discuss more fully a number of variations on
the neutrino lasing phenomenon.

\section{BOLTZMANN EQUATIONS} 	\label{sec:boltzmann}

Allowing only processes (\ref{eq:reactions})  and their inverse decays, we wish
to develop
the formalism that will allow us to follow the evolution of the
occupation number
distributions $\ftau,\ \ff,\ \fb$.
Following Wagoner (1979),
 the net rate at which H's of energy $\Etau$ are being created
from $\F$'s and B's in momentum space elements $d^3 \pf$, $d^3
\pb$ (i.e. the inverse decay rate minus the decay rate)
is

\begin{equation}		\label{eq:d2fA}
d^6 \dot \ftau = - (\ftau (1 - \ff) (1 + \fb) - \fb \ff (1-\ftau))
{ W(k_H \rightarrow k_F, k_B) \over \Etau}{d^3 \pf\over \Ef} {d^3 \pb\over
\Eb}
\end{equation}
where $k$ is the four momentum, and $W$ is the invariant transition rate
as defined by Wagoner.
Here, however, the occupation numbers are isotropic functions of the 3-momentum
so the rate at which H's of energy $\Etau$ are being created from $\F$'s and
$\B$'s  in $\Ef;\;\Ef + d\Ef$, $\Eb;\;\Eb + d\Eb$ is
\begin{equation}    \label{eq:d3fA}
d^2 \dot \ftau = -(\ftau (1 - \ff) - \fb (\ff - \ftau))
\Etau^{-1} F(\Etau, \Ef, \Eb) \pf \pb d\Ef d\Eb
\end{equation}
where
\begin{equation}
F(\Etau, \Ef, \Eb) \equiv {1\over 4\pi} \int d\Omega_F d\Omega_B d\Omega_H
W(k_H \rightarrow k_F, k_B)
\end{equation}
and the integrations are over directions of momenta satisfying
$p_i^2 = E_i^2 - m_i^2$.

Given a specific form for the H-F-B interaction (e.g.\ Yukawa type)
one could  directly calculate the transition rate $W$ and then do
the integration over directions.
It is simpler to obtain F from simple
relativistic kinematics of the decay process. This
is sufficient to determine F up to two parameters. One of these
can be taken to be $\Gamma_0$, the free decay rate, and the other
parameter $\alpha$ is related to the spin of the H's and describes the
anisotropy of the decays in the H rest frame.

\subsection{Kinematics}					\label{sec:kinematics}

In the rest frame of the
decaying particle (primed quantities), the decay products carry equal and
opposite momenta
$|\pb'| = |\pf'| = p_0$
and energies $E'_i = \sqrt{{p'}_i^2 + m_i^2}$, where $i$ = F,B.
With $\Ef' + \Eb' = \Eh' = \mh$
we find $p_0 = m_0/2$ where
\begin{equation}						\label{eq:m0definition}
m_0^2 = \mh^2 - 2(m_B^2 + m_F^2) + (m_B^2 - m_F^2)^2/\mh^2
\end{equation}
For $\mh \gg \mf,\mb$, the situation of most interest here,
$p_0 \simeq \mh / 2$.
Lorentz transforming to the cosmic frame, we find that for a H with
energy $\Eh = \gamma \mh$, the energy of the decay products is
\begin{equation}  			\label{eq:eofmu}
E_i = \gamma(E'_i + \beta \mu p'_i)
\end{equation}
where  $\beta = |{\bf v}| / c =(1 - \gamma^{-2})^{1/2}$ and
$\mu$ is the cosine of the angle in the rest frame between
the velocity of the decay product and
the direction of the parent velocity in the cosmic frame.
The maximum and minimum values of the B and F energies are therefore
\begin{equation}
E_i^\pm = \gamma(E'_i \pm \beta p'_i)
\end{equation}
or, in terms of $\Etau$ and the particle masses,
\begin{equation}		\label{eq:eilimits}
E_i^\pm = (\mo / 2 \mtau) (\Etau\sqrt{1 + 4 m_i^2 / \mo^2} \pm \ph)
\end{equation}
and
the corresponding maximum and minimum $\Etau$ for given decay product energy
$E_i$ is
\begin{equation}		\label{eq:ehlimits}
\Etau^\pm(E_i) = {\mo \mt \over 2 m_i^2}
	\Bigl[E_i{\sqrt{1+{4m_i^2\over \mo^2}}} \pm {\sqrt{E_i^2-m_i^2}} \Bigr]
\end{equation}
These kinematic limits on the allowed combinations of $\Eh,\Eb$ are
shown in figure 1:

Figure~1 is crucial for understanding neutrino lasing.
The limits (\ref{eq:eilimits}) should be thought of as
providing an additional constraint on the energy conserving plane $\Eh = \Eb +
\Ef$
in 3-dimensional ($\Eh,\Eb,\Ef$) energy space.
In figure~1 we see the projection of this accessible
energy surface on the $\Eh,\Eb$ plane.
For $\ph = 0$, the boson
energy is just $\Eb = \sqrt{\mb^2 + m_0^2 / 4}$
and is therefore approximately $\mh/2$ for $\mh \gg \mb,\mf$.
For $\ph > 0$ there is a range of energies, corresponding to the
range of angles $\mu$ for the emission of the decay products.
Of particular interest here is the lower bound on the boson energy
(realised for bosons emitted in the backwards direction), which,
for $\mb \ll \mh$ is
\begin{equation}
\Eb^- \simeq {4\mb^2\Eh^2 + \mh^4\over 4\mh^2\Eh}.
\end{equation}
This reaches a minimum at
\begin{equation}				\label{eq:estardefinition}
\Estar \equiv {\mh^2 \over 2 \mb} \Bigl(1 + {\mb^2-m_F^2\over \mh^2}\Bigr)
       \simeq {\mh^2 \over 2 \mb}.
\end{equation}
For $\mh \ll \Eh \ll \Estar$, we have $\Eb^- \simeq \mh^2 / 4 \Eh \ll \Eh$.
In this regime the backwards emitted boson appears backwards moving in the
cosmic frame. For $\Eh = \Estar$ the backwards emitted boson appears at
rest in the cosmic frame, and for $\Eh > \Estar$, $\Eb^- \simeq (\mb/\mh)^2
\Eh$,
and the backwards emitted boson moves forwards in the cosmic frame.
Again we have $\Eb^- \ll \Eh$ and, except for extremely high energies
$\Eh \gg \mh (\mh/ \mb)^2$, we have $\Eb \ll \mh$.

\subsection{Spin Polarisation}					\label{sec:spin}

The decay product energy distribution (for free decays) depends, through
(\ref{eq:eofmu}), on the distribution of $\mu$,
the decay product momentum direction in the H-rest frame.
In KMS, we assumed that the decays were isotropic: $P(\mu) = 1/2$.
However, as the heavy neutrinos are
spin polarised this is not necessarily correct.
Massless neutrinos are purely left handed (in chirality and helicity).
Since we are interested in neutrinos with masses of a few keV or less,
we might expect the H's to be essentially
massless when the weak interactions decouple at $T\simeq 1\MeV \gg \mh$
and that the H's should therefore be essentially 100\% polarised
with spin antiparallel
to their direction of motion in the cosmic frame.
The distribution $P(\mu)$  depends on the details of the
H-F-B interaction (see \S\ref{sec:majoron}),
but, for fully polarised H's is given
quite generally (see e.g.
discussion in Feynman \etal, 1965 \S17-5) by
\begin{equation}  \label{eq:pofmu}
P(\mu) = {1\over2}(1-\alpha\mu)
\end{equation}
where the model dependent parameter $\alpha$, lies in the range
$-1\leq\alpha\leq1$.
The probability distribution of boson energies for a particular $E_H$
is then, from (\ref{eq:pofmu}, \ref{eq:eofmu}), a linear ramp:
\begin{equation}				\label{eq:pofe}
P(E_B) =
\left\{
\begin{array}{ll}
{(E_B^+ - E_B^-) + \alpha(E_B^++E_B^-)
- 2\alpha E_B\over (E_B^+-E_B^-)^2}, &
\mbox{if $E_B^-\leq E_B\leq E_B^+$} \\
0 & \mbox{otherwise}
\end{array}
\right.
\end{equation}

Now, by definition,
\begin{equation}
d^2\dot\fh = \dot\fh P(\Eb, \Ef) d\Eb d\Ef
\end{equation}
but as energy is conserved in the decays, $P(\Eb,\Eh) = P(\Eb) \delta(\Ef + \Eb
- \Eh)$,
and as the net rate of decays is just the time dilated free decay rate:
$\dot\fh = - (\mh/\Eh) \Gamma_0\fh \equiv \Gamma_d \fh$, we find
\begin{equation}		\label{eq:d2fB}
d^2 \dot \ft(\Et) = -{\mh\over \Et } \Gamma_0\ft(\Et) \delta(\Ef + \Eb - \Et)
P(E_B) d\Eb d\Ef
\end{equation}
All this is for unpopulated final states ($\ff = \fb = 0$).
It follows from elementary field-theoretic considerations that
to allow for the finite occupation number for the decay products
we must replace $\fh$ here by $\fh(1-\ff)(1+\fb) - \ff\fb(1-\fh)$,
where the first term now includes the Pauli blocking factor $1-\ff$ and
the stimulation term $1+\fb$, and the second term is for
inverse decay processes.
Thus we have
\begin{equation}		\label{eq:d2fC}
d^2 \dot \ft(\Et) = -{\mh\over \Et } \Gamma_0
(\ft(1-\ff)(1+\fb) - \ff\fb(1-\fh))
\delta(\Ef + \Eb - \Et)
P(E_B) d\Eb d\Ef.
\end{equation}
Comparing (\ref{eq:d3fA}) with (\ref{eq:d2fC}) we see that
these are the same if we make the identification
\begin{equation}
F(\Et, \Ef, \Eb) = {\mt \Gamma_0\delta(\Ef + \Eb - \Et) P(\Eb)
\over \pb \pf }
\end{equation}

Finally, integrating (\ref{eq:d2fC}) over $\Eb,\ \Ef$,   we obtain
\begin{equation}
\dot \ftau(\Etau)  = {- \mt^2 \Gamma_0 \over \mo \Etau \pt}
\int\limits_{\Eb^-(\Etau)}^{\Eb^+(\Etau)} d\Eb
(\ftau (1 - \ff) - \fb (\ff - \ftau))
\biggl[ 1 + \alpha{\Eb^++\Eb^--2\Eb \over \pt \mo/\mtau}\biggr]
\end{equation}
where again $\ff = \ff(\Etau - \Eb)$.
This is a convenient point to make the transition to an expanding
universe: if we let $E = E'/a$, $p = p'/a$ and drop the primes, the only
changes
are that we must replace $m$ by $m a$ and we must replace $\dot
\ftau$ by the time derivative at constant comoving momentum $\dot f \rightarrow
\partial f / \partial t + (H m^2 a^2 / E) \partial f / \partial E$, since in
the
absence of reactions it is the occupation number at fixed comoving momentum
that is adiabatically conserved.

The net rates for formation of $\B$'s and $\F$'s are obtained in a similar
fashion and we obtain the final form for the Boltzmann equations:
\begin{mathletters}			\label{eq:boltzmann}
\begin{eqnarray}
{\partial \ftau \over \partial t} + {H \mt^2 a^2 \over \Etau}
{\partial \ftau \over \partial \Etau} & =
{- \mtau^2 a \Gamma_0 \over  \mo \Etau \ph}\int
\limits_{\Eb^-(\Etau)}^{\Eb^+(\Etau)}
d\Eb S(\Etau, \Eb, \Etau - \Eb)
 A(\Eh,\Eb) \\
{\partial \fb \over \partial t} + {H \mb^2 a^2 \over \Eb}
{\partial \fb \over \partial \Eb} & =
 {\mtau^2 h_B a \Gamma_0 \over  \mo \Eb \pb}
\int \limits_{\Etau^-(\Eb)}^{\Etau^+(\Eb)} d\Etau
S(\Etau, \Eb, \Etau - \Eb)
 A(\Eh,\Eb) \\
{\partial \ff \over \partial t} + {H \mf^2 a^2 \over \Ef}
{\partial \ff \over \partial \Ef} & =
{\mtau^2 a \Gamma_0 \over  \mo \Ef \pf}
\int \limits_{\Etau^-(\Ef)}^{\Etau^+(\Ef)} d\Etau
S(\Etau, \Etau - \Ef, \Ef) A(\Eh,\Eh - \Ef)
\end{eqnarray}
\end{mathletters}
where
\begin{equation}
S(\Etau,\Eb,\Ef) = \ftau(\Etau) (1 - \ff(\Ef))(1 + \fb(\Eb)) -
(1 - \ftau(\Etau)) \ff(\Ef)\fb(\Eb),
\end{equation}
\begin{equation}
A(\Eh,\Eb) = \Bigl[ 1 + \alpha {\Eb^+(\Eh) + \Eb^-(\Eh) - 2\Eb\over\ph
\mo/\mtau}\Bigr],
\end{equation}
and $h_B = 2 (1)$ if $\Bbar = (\ne) \B$.
These provide a complete set of coupled equations which
allow us to evolve the distribution functions
$\ftau(E, t),\ \ff(E, t),\ \fb(E, t)$.

We emphasise that these equations include only the process
(\ref{eq:reactions}).
In particular they do not include the HB, HF and BF scattering processes that
are
the automatic consequences of the decay process
at higher order in the BFH coupling, nor do they include
number changing weak interactions.
The latter are important at early times and the former may potentially
become important at late times, and if so
the collision terms developed here would need to be augmented.

\subsection{Limiting Cases} 		\label{sec:limitingcases}

\subsubsection{Free decay} 		\label{sec:freedecay}

Consider the case that initially $\ff = \fb = 0$ so that $S = \fh(\Eh)$
is independent of $\Eb$ and may be taken out of the integral in the
first of equations (\ref{eq:boltzmann}).
Since from (\ref{eq:eilimits}) $\Eb^+ - \Eb^- =
\mo \ph / \mtau$ the integral is trivial and we obtain
\begin{equation}
\dot \ftau = - (\mtau \Gamma_0 / \Etau) \ftau
\end{equation}
so the particles simply decay at the free decay rate $\Gamma_0$ with a
$1/\gamma_H$
modification for time dilation.
These decays occur at a
temperature $T_0$ when the Hubble parameter, $H(T_0) \equiv \Gamma_0$.
As discussed in the introduction, this result is really only
valid for non-relativistic decays (in which case $\ff$ and $\fb$
remain small) and in the relativistic case one must allow for the
build up of decay products.

\subsubsection{Thermalisation} 		\label{sec:thermalisation}

Another simple but interesting solution of (\ref{eq:boltzmann}) is the thermal
distribution:
\begin{equation}		\label{eq:thermalf}
f_i = (\exp(\beta( E_i-\mu_i)) \pm 1)^{-1}
\end{equation}
with the plus (minus) sign for fermions (bosons).
Here $\beta = 1/T$ is the inverse temperature and
$\mu_i$ is the chemical potential for the $i$th species.
The collision terms in (\ref{eq:boltzmann}) vanish provided
\begin{equation}		\label{eq:chempot}
\mu_H = \mu_F + \mu_B.
\end{equation}
For the boson we must have $\mu_B < \mb$.
As $\mu_B$ tends to $\mb$ from
below, the number of zero momentum particles becomes large
and one has a Bose condensate.
For $m \ll T$,
the second (\lq\lq{}cosmological\rq\rq) term in the lhs of equations
(\ref{eq:boltzmann})
is also negligible and
consequently the thermal distributions with constant $\alpha_i\equiv\mu_i/T$
will be valid solutions.

If we start with some arbitrary
non-equilibrium state and assuming that equilibrium is reached
then if we
specify the initial number and energy density in each
particle species,
the final parameters  $\beta$, $\alpha_{\tau}, \af$ and $\ab$ (or $n_0$)
are determined from the 3 continuity equations
\begin{mathletters}
\begin{eqnarray}
\nh + \nf & = {\nh}_i + {\nf}_i \\
\rh + \rf + \rb/h_B & = {\rh}_i + {\rf}_i + {\rb}_i/h_B \\
\nh + \nb/h_B & = {\nh}_i + {\nb}_i/h_B
\end{eqnarray}
\end{mathletters}

\medskip\noindent
together with the algebraic constraint (\ref{eq:chempot}).
Here $n = n_{\rm phys} a^3$ and $\rho = \rho_{\rm phys} a^4$ are
suitably scaled quantities which remain constant during the
expansion of the universe. This, in essence, is the calculation of
 Madsen (1992).  Assuming that
the H's initially had a thermal distribution with $\ah = 0$; that
the B's were absent and that the F's were  also
absent  (or had $\ff = \fh$), he concluded that
Bose condensation formed for a large range of initial conditions.

There are several question marks over this result:

\noindent
i) For a true Bose condensate to form one must be able to access
zero momentum bosons.  Since only the processes (\ref{eq:reactions}) are
assumed to
 be operative, the H's must be able to decay and produce a  zero-momentum
boson. From (\ref{eq:eilimits}) we see that this is
kinematically possible, but only from a specific H-energy: $\Eh = \Estar$.
In order to get a substantial condensate to form it is necessary
that there be H's with this energy (they would be exponentially
suppressed for $T \ll \Estar$ for instance) and that they get replenished
as the decays proceed.  Certainly the assumption that the
timescale for equilibration is just the time dilated free decay
time seems questionable.

\noindent
ii) For $T \sim \mh$ one must allow for the second term
on the lhs of (\ref{eq:boltzmann}) which invalidates the constant $\beta$,
$\alpha$
result.  This is a pity since for F being a known neutrino,
Madsen obtained
a Bose condensate only for thermalisation happening at around
$T = \mh$.  One could readily allow for this by developing
a dynamical set of equations that allow one to proceed through
a sequence of thermal states.  However, this is of dubious
merit since for $T \sim \mh$ (and $\mh \gg \mb$) one automatically
has $T \ll \Estar$, so the decay to zero momentum bosons
is suppressed.

\noindent
iii) It is not clear that one should assume that
the B's and the F's initially  are completely absent.
The physical idea here is
that these particles were decoupled by $\TQCD$.  If they were
coupled before this then their abundance would be smaller than the
thermal equilibrium abundance by a factor of $\eta_{QCD}^{3}\simeq1/6$ (though
if they were never in equilibrium then they might well have effectively
zero abundance).  Including this initial abundance reduces the
final cold:hot ratio.

\noindent
iv) The calculation is
incomplete since not all of the H's have decayed and the question
of what happens to the remaining H's --- which must eventually decay ---
is left unanswered.

The proper treatment of these effects,
and an examination of the role that spin
polarisation plays in this process,
is the focus of this work.

\section{LASING AND QUASI-CONDENSATES} \label{sec:lasing}
\subsection{Lasing: Production of the Quasi-Condensate}

We now try to address some of the above issues, focusing on the production of
low momentum bosons.
At first, we keep the discussion general, without identifying
the H,F or B with any known particles, and simply
assuming that we are provided with some arbitrary initial occupation numbers;
deferring until later the question of how these initial conditions might
arise.

As we have shown elsewhere (KMS), if there is initially
an excess of H's over F's this can drive a process of runaway stimulated
decays or lasing.
The basic physics can be understood quite simply.  Consider
for the moment what would happen if we started with a thermal gas
of H's with $T \gg \Estar$ and no F's or B's, and also consider for simplicity
isotropic decays ($\alpha = 0$).  From (\ref{eq:pofe}) we see that
the H's at a given energy $\Eh$ (typically
of order $T$) will start to decay into B's uniformly distributed
in energy on the interval $\Eb^-\leq \Eb\leq \Eb^+$.
The minimum $\Eb$  for $\mb \ll \mh$ is
\begin{equation}
\Eb^- \simeq {4\mb^2\Eh^2 + \mh^4\over 4\mh^2\Eh}.
\end{equation}
Now, as we have discussed,  a key feature of relativistic decays is that,
for $\Eh \geq \Estar$,
$\Eb^- \sim (\mb/\mh)^2 \Eh$ which is
much smaller than $\Eh$; these correspond to
decays in which the B is emitted in the backward direction.

While the fraction of low energy B's is initially quite
small, the phase space at low energy is
smaller than the phase space initially occupied by the H's
by a factor $\sim (\pb/\ph)^3 \simeq (\Eb/\Eh)^3$.
Since  the energies of the B's are
uniformly distributed in $\Eb^-\leq\Eb\leq\Eb^+$,
a fraction ${\cal F} \simeq  (\Eb-\Eb^-)/(\Eb^+-\Eb^-) \simeq \Eb^-/\Eh$
will have energy $\Eb\sim\Eb^-$.
Therefore,  once $(\Eb/\Eh)^2$ of the H's
have decayed, the boson occupation number at $\Eb \sim \Eb^-(T)$ will
be of order unity even if they were totally absent initially.  This initiates
a runaway process and very soon essentially all of the decays will
be into low momentum B's.
This does not create a true $p=0$ Bose condensate; we refer to the
low momentum B's with $\fb \gg 1$ as a quasi-condensate.

The analogy with an ordinary laser is made complete if we think
of the H and F as the excited and ground states of a two-state system
with the transition involving the emission or absorption of a B.
Just as with an ordinary laser, the process saturates when $\fh = \ff$ ---
when the interaction is equally likely to excite an F
(creating an H and giving up B-energy)
as to stimulate a decay.
Clearly, some non-equilibrium process is necessary in order to create the
initial population inversion, $\fh > \ff$, since in equilibrium
the abundance of the more massive state will always be
equal to or smaller than that of the lower mass state.

Let us now investigate this process in detail.
The goal here is to quantify the rate at which the lasing proceeds,
and also to identify the momentum of the bosons produced. This expands on
KMS in that we explicitly account for the effects
of spin polarisation. We will also generalise the picture
to the case where there is more than one light decay channel.

We first need to simplify the Boltzmann equations (\ref{eq:boltzmann}).
Consider the equation for $\dot\fh$.
Since $f \le 1$ for the fermions and the term in square
brackets is $< 2$ it is easy to see that
for the $\fb$-independent terms there is a natural timescale
which is just the time dilated free-decay time,
so these terms, which represent ordinary H decay and inverse decay,
 are ineffective for $T \gg T_d$.
The terms involving $\fb(\ftau - \ff)$ can, however,
be effective even when $T \gg T_d$, if $\fb$ is sufficiently large.
Consider, now the equation for $\dot\fb$.
Since we are looking for  solutions with $\fb\gg1$
we ignore those terms in the $\fb $ equation which are independent of $\fb$.
(Since they are strictly positive this is a conservative approximation.)
Then from (\ref{eq:boltzmann}) we have
\begin{equation}		\label{eq:fbdot}
\dot \fb \simeq	 {h_B \Gamma_0 \mt^2 \fb\over \mo \Eb \pb}
\int\limits_{\Eh^-(\Eb)}^{\Eh^+(\Eb)} d\Etau
(\ftau(\Etau) - \ff(\Etau - \Eb))\Bigl[ 1 + \alpha{\Eb^+ + \Eb^- - 2\Eb\over
						  \ph m_o/\mh}\Bigr]
\end{equation}
This is of the form
\begin{equation}
\dot \fb = \Gammalase \fb
\end{equation}
where $\Gammalase $ is some function of $\Eb$, $\fh$ etc.\ and
indicates that $\fb$ may grow (or decay) exponentially.
As we are interested in low momentum bosons: $\Eb \ll \Ef,\Eh$, we
can safely approximate $\ftau(E) - \ff(E - \Eb)$ by
$\ftau(E) - \ff(E)$, and we see that a necessary condition for
exponential growth is $\fh > \ff$.

Assuming this condition is met,
what does $\Gammalase(\Eb)$ look like? In particular, we
would like to identify the energy $\Eb$ for which $e$-folding rate is
maximised. For concreteness, consider the case that the F's have a thermal
distribution but with $T_F < T_H$, say $T_F = T_H / 2$ (this is
roughly the relevant case if the F decoupled prior to
quark confinement).  The factor $\fh - \ff$ will now be a bell shaped
curve with centre at $E \sim T$ and with width $\delta E \sim T$.

It is now easy to see from figure 1
that for $\alpha = 0$, the
integral in (\ref{eq:fbdot}) will be some kind of sigmoid curve:
for $\Eb \ll \Eb^-(T)$,
the upper limit of the $\Eh$ integral is $\Eh^+(\Eb) \ll T$,
so the integrand is effectively zero over the complete range of integration;
while for $\Eb \gsim \Eb^-(T)$ the integral is of order $T$.
For $\Eb \gg \Eb^+(T)$,
the integral will fall again, but this is not relevant here.
(For a very narrow $\Eh$ distribution the sigmoid would become a step,
reflecting the sharp lower edge to the distribution (\ref{eq:pofe}),
and the width in the realistic case stems from
the finite range of $\Eh$ in the thermal distribution.)
Multiplying the sigmoid by the prefactor $1/\Eb\pb$
(and noting that since $\Eh\sim T\gg\Estar$ therefore
$\Eb^-\sim (\mb/\mh)^2T\gg\mb$ and $\pb^-\simeq\Eb^-$),
we see that $\Gammalase$ will peak around $\Eb \simeq \Eb^-(T)$
with value
\begin{equation}		\label{eq:gammalase}
\Gammalase \sim { \mh T  \over (\Eb^-)^2}\Gamma_0
\sim \Gamma_0\cases{
{\mh^5  / \mb^4 T} & $ T \gg \Estar $ \cr
{T^3  / \mh^3 } & $  T \ll \Estar $ \cr
}
\end{equation}
and the boson momentum for which this maximum rate is realised is
$\pb \sim \Eb^-$:
\begin{equation}
\pb
\sim \cases{
{(\mb^2/ \mh^2) T} & $  T \gg \Estar$  \cr
{\mh^2 / T } & $  T \ll \Estar $ \cr
}
\end{equation}
Comparing (\ref{eq:gammalase}) with the free decay rate:
\begin{equation}
\Gamma_D
\sim \cases{
(\mh / T) \Gamma_0 & $  T \gg \mh $ \cr
\Gamma_0 & $  T \ll \mh $  \cr
}
\end{equation}
we see that $\Gammalase \gg \Gamma_D$.

A necessary condition for lasing to be effective is that
$\Gammalase > H$.  From (\ref{eq:gammalase}) we see that
$\Gammalase / H \propto 1/T^3$ for $ T \gg \Estar$.
(We assume throughout that we are in the radiation dominated epoch.)
Thus, at sufficiently
early times lasing will be inoperative, but will switch on
when $\Gammalase \sim H$.
For $T \ll \Estar$,
$\Gammalase / H \propto T$.
What this means in practice is that is that either
lasing will become effective before the temperature falls to
$\Estar$, or will never be effective (though see below for $\alpha = -1$).

We now generalise this to arbitrary $\alpha$, in which case
we need to consider the square bracketed term in (\ref{eq:fbdot}).
The result only changes qualitatively if $\alpha = -1$ or is very close to
$-1$.
In that case
the integral is still essentially zero for $\Eb \ll \Eb^-(T)$, but
now ramps up linearly:
\begin{equation}
\int d\Eh\ldots \sim T{\Eb-\Eb^-(T)\over\Eb^+(T)-\Eb^-(T)} \sim (\Eb-\Eb^-(T)).
\end{equation}
Multiplying by the prefactor $1/\Eb \pb$, we see that
the maximum growth rate is again at $\Eb - \Eb^-(T) \simeq \Eb^-(T)$,
but is now smaller
by a factor $\sim \Eb^-(T) / T$
and the maximum growth rate is
\begin{equation}
\Gammalase \sim { \mh \Gamma_0 \over \Eb^-(T)}
\sim
\cases{
{\mh^3 \Gamma_0 / \mb^2 T} & $ T \gg \Estar $ \cr
{T \Gamma_0 / \mh} & $ T \ll \Estar $ \cr
}
\quad\quad\quad\quad \alpha = -1
\end{equation}

For $T \gg \Estar$
the lasing rate grows with time ($\Gammalase \propto 1/T$) while the expansion
rate $H = \Gamma_0 (T / T_0)^2$ decreases, and they will become equal at
\begin{equation}
\Tlase^3 = T_0^2 \mh (\mh/\mb)^n
\end{equation}
with $n = 2(4)$ for $\alpha = (\ne) -1$.
Here $T_0$ is defined such that $H = \Gamma_0(T/T_0)^2$.
(If $T_0<\mh$, then this is the value $H$ would have had
if we ignored the fact that the energy density of the universe
would have become $\mh$ dominated at $T\sim\mh$.)  We will also
express things in terms of the more physical $T_d$, which includes the
effects of time dilation, $H(T_d) \equiv \Gamma_d$.
The temperatures $T_d$ and
$T_0$ are related by $T_d={\rm min}(T_0,(\mh T_0^2)^{1/3})$.

The condition that $\Tlase \gg \Estar$ is therefore equivalent
to the following limit on the free-decay temperature
\begin{equation}
T_0 / \mh \gg
\cases{ (\mb/\mh)^{1/2}  & $\alpha\ne -1$ \cr
	      (\mb/\mh)^{-1/2} & $\alpha=-1$\cr}
\end{equation}
or equivalently a limit on the time-dilated free decay temperature
\begin{equation}
T_d / \mh \gg
\cases{ (\mb/\mh)^{1/2}  & $\alpha\ne -1$ \cr
	      (\mb/\mh)^{-1/3} & $\alpha=-1$.\cr}
\end{equation}

Below $T = \Estar$,  $\Eb^-(T)\sim \mh^2/T$,
so that
\begin{equation}
\Gammalase(T\ll\Estar) \simeq \cases{ \Gamma_0 (T/\mh)^3 & $\alpha\ne-1$ \cr
			              \Gamma_0 (T/\mh) & $\alpha\simeq-1$ \cr}
\end{equation}
For $\alpha \ne -1$, the ratio $\Gammalase / H$ decreases with
time, so, as noted above,
 if lasing does not occur by $T \simeq \Estar$ it never will.
For $\alpha \simeq -1$, lasing will occur at
\begin{equation}
\Tlase(\alpha\simeq-1, T_0\ll \mh^{3/2}\mb^{-1/2}) = T_0^2/\mh,
\end{equation}
so long as $T_0\gg\mh$.

\subsection{Post-Lasing  Reprocessing}

Assuming lasing occurs at some $T \gg \Estar$ then we still need to
ask what happens subsequently.
The bosons are produced with energy just above the minimum value $\Eb^-(T)$
for typical H energy $\Eh \sim T$.  After the lasing saturates, the
bosons will be in extremely tight contact with the H's and F's;
the net rate for transitions is small, but the total rate is enormous.
As the populations redshift, the bosons will remain in contact since
$\Eb^-$ is just a multiple of $T$, until $T \simeq \Estar$
but below this temperature contact will be lost (as $\Eb^-/T$ then
increases like $\mh^2 / T^2$ --- see eq.(\ref{eq:eilimits})) and the
cold bosons freeze out never to communicate again with the fermions.
However, we have assumed all along that $T \gg T_d$.  What
if $T_d > \Estar$? In this case the cold laser-produced bosons are still
in tight contact, but decays coupling the fermions to bosons
with energy $\Eb \sim T$ now become active at $T \simeq T_d$.
The approximations used above break down, but it is reasonable
to expect that the result will be very similar to the equilibration
calculation of Madsen, but with a quasi-condensate rather than
a true condensate.  However, as we now need to populate the $\Eb \sim
T$ states, this necessarily means that some of the cold bosons must
be reabsorbed and the cold fraction falls.

In either case the populations stabilise by $T = \Estar$ when the
H's are still relativistic.  For $T_d < \Estar$, we have
$\fh = \ff$ with grey body Fermi-Dirac spectra and for $T_d > \Estar$
we have thermal distributions but with $\ah = \af$.
The remaining H's will decay when $T \sim \mh$ when they become
transrelativistic if $T_0 > \mh$. If $T_0 < \mh$, which is only
compatible with lasing if $\alpha \ne -1$, then the final decays
occur at $T \simeq T_0$ and produce very hot B's and F's.

If $\alpha\simeq-1$ and $\Tlase \leq\Estar$, the cold laser-produced bosons
are quickly redshifted out of contact and we have
$\fh = \ff$ with grey body Fermi-Dirac spectra.

We can therefore distinguish 4 cases:
\begin{itemize}
\item[case A:] $T_d > \Estar$ for any $\alpha$
\begin{enumerate}
\item lasing occurs at $\Tlase \gg T_d$ giving grey-body
fermion spectra and quasi-condensate (plus any initial warm
bosons).
\item thermalisation occurs at $T_d$ while cold
bosons are still tightly coupled giving essentially
thermal distributions aside from
quasi-condensate in place of true Bose condensate, and
with abundances given by Madsen's calculation (suitably modified if necessary
to allow for finite initial F and B temperature).
\item cold bosons freeze out at $T \simeq \Estar$
\item final decays occur at $T \sim \mh$ to make hot
bosons.
\end{enumerate}
\item[case B:] $\Estar > T_d > \mh$ for $\alpha \ne -1$ or
$\Estar > T_d > \mh (\mh/\mb)^{1/3}$ for $\alpha = -1$
\begin{enumerate}
\item lasing occurs at $\Tlase \gg \Estar$ as in A
\item cold bosons freeze out at $T \simeq \Estar$
\item at $T_d$, processes linking particles with $p \sim T$
become effective, and alter the B,F and H distributions
for momenta $p \sim T$.
\item final decays occur at $T \sim \mh$ out of equilibrium to make hot
bosons
\end{enumerate}
\item[case C:] $\mh(\mh/\mb)^{1/3}> T_d > \mh$ for $\alpha\simeq-1$
\begin{enumerate}
\item lasing occurs at $\Tlase > T_d$ as in A, B
\item no thermalisation
\item final decays occur at $T \sim T_d$ out of equilibrium
to make bosons with $T_B \sim T_H$
\end{enumerate}
\item[case D:] $\mh > T_d > \sqrt{\mb\mh}$ for $\alpha \ne -1$
\begin{enumerate}
\item lasing occurs at $\Tlase \gg T_d$ as in A, B
\item no thermalisation
\item final decays occur at $T \sim T_d$ out of equilibrium
to make very hot bosons
\end{enumerate}
\end{itemize}

These various regimes of ($\mh,T_d$) parameter space are shown in figure~2, for
the cases
(a) $\alpha\neq-1$ and (b) $\alpha=-1$.
The upper limits $T_{lase} < 2.3$MeV,
applies only in the case where H is still coupled after the
QCD phase transition,
for example H$=\nu_\tau$ (for further discussion see $\S 5$).
It follows from the requirement that  the
that the helium-4 mass fraction
produced in standard big bang nucleosynthesis remain less than
the inferred primordial fraction $Y_P\leq 0.24$
(see \eg~ Pagel  {\it et al.} 1992).  For a recent discussion of Standard Big
Bang Nucleosynthesis see
Smith, Kawano and Malaney (1993).  Any new particle  (such as the B)
must satisfy the condition that it contribute to the energy density  at
$2.3$MeV less than
$0.3$ effective neutrino families ($\Delta N_\nu \le 0.3$) (\eg~ Walker \etal~
1991).
Any particle
not thermally coupled at quark confinement would contribute less than $\sim
0.1$
to $\Delta N_\nu$ after completion of the phase transition.

The cold fraction can be calculated analytically in cases B, C, and D
(and without recourse to full blown Boltzmann equations
in A) as  follows:
Let's assume that prior to lasing the three species all have
zero-chemical potential thermal distributions, but with
the temperature of the B's and F's suppressed by a factor $\eta^{1/3}$ (so
their
numbers are suppressed by a factor $\eta$ relative to the
abundance they would have had if they had had $T = T_H$).
The factor $\eta$ would be $\simeq 1/6$ if, for example, the B's and F's
had decoupled at some temperature not too far above the quark confinement
temperature, but, as
discussed, it may be that these particles were never in equilibrium
or decoupled at some extremely early time, in which case $\eta$ would be
effectively zero.

Prior to lasing then, the abundance of the particles (summing over
spin and particle/antiparticle states):
\begin{equation}
\nb^C = 0; \quad \nb^H = 4\eta / 3 h; \quad \nh = 1; \quad \nf = \eta
\end{equation}
where the superscripts H,C refer to hot and cold.
After lasing, some fraction $d$ of the H's will have decayed and we have
\begin{equation}
\nb^C = d; \quad \nb^H = 4\eta / 3 h; \quad \nh = 1-d; \quad \nf = \eta + d
\end{equation}
but the end result of lasing is to equalise $\nh$ and $\nf$, so we
have $d = (1 - \eta) / 2$
and we find for the cold fraction
\begin{equation}
{n_B^C\over \nb^C + n_B^H} = {(1-\eta)\over   2(1+4\eta/3h)}
\end{equation}

A potential problem with the foregoing model is the low yield of cold
particles, $n_B^C/n_B^H \leq 1$.
A simple modification is to increase
the number of decay channels by increasing the number
of light fermions into which the H can decay, and we find for $N$ decay
channels
\begin{equation}
{n_B^C\over \nb^C + n_B^H} = {N(1-\eta)\over (1+N)(1+4\eta/3h)}
\end{equation}
where again we have taken $T_B=T_{Fi}=\eta^{1/3}T_H$ for $i=1,...N$
initially.
In case A we expect a modified analytic result for the cold fraction.

\section{Neutrino-Lasing in the Context of the Singlet-Majoron Model}
\label{sec:majoron}

In the singlet-majoron Model (Chikashige, Mohapatra and Peccei 1981) a new
Higgs singlet field $S$ is
introduced with a non-zero expectation value $<S>_o=u/\sqrt 2$ which
spontaneously
violates lepton number:
\begin{equation}
S={1\over \sqrt 2} (u + \rho + i\chi)
\end{equation}
where $\rho$ is a Higgs scalar and $\chi$ is the majoron --
 the massless degree of freedom associated with the
$U(1)$ symmetry.
In the context of this model,
the most general description for the mass and Yukawa coupling
 terms of the neutrino Lagrangian
is
\begin{equation}
{\cal L}=\sum_{\alpha\beta}\bigg\{
M_R^{\alpha\beta}\overline\nu_{\alpha(R)}\nu_{\beta(R)}
+m_D^{\alpha\beta}[\overline\nu_{\alpha(L)}\nu_{\beta(R)}
+\overline\nu_{\alpha(R)}\nu_{\beta(L)}] +
 {i\over \sqrt
2}h^{\alpha\beta}{\chi\overline\nu_{\alpha(R)}\nu_{\beta(R)}}\bigg\}\>,
\end{equation}
where $\nu_\alpha$ are Majorana fields, which can be described in terms
of four component spinors $\psi_\alpha$ as
\begin{eqnarray}
\nu_{\alpha(L)}=&{1\over\sqrt{2}}(\psi_{\alpha L}
+(\psi_{\alpha L})^c) \nonumber \\
\nu_{\alpha(R)}=&{1\over\sqrt{2}}(\psi_{\alpha R}
+(\psi_{\alpha R})^c)\>
\end{eqnarray}
As far as the weak interactions are concerned, $\nu_{\alpha(L)}$ are active
eigenstates, and $\nu_{\alpha(R)}$ are sterile states.
The mass terms $M_R^{\alpha\beta}$ arise from Yukawa couplings
to the singlet field $S$:
\begin{equation}
M_R^{\alpha\beta}=h^{\alpha\beta}u\>,
\end{equation}
whereas the Dirac mass terms $m_D^{\alpha\beta}$
arise from couplings to the normal Higgs doublet,
with non-zero expectation value $<H>=v/\sqrt 2$:
\begin{equation}
m_D^{\alpha\beta}=g^{\alpha\beta}v\>.
\end{equation}

Given $n$ neutrino families, the
generic form for the  mass matrix of $2n$ Majorana fields can be written
\begin{equation}		\label{eq:massmatrix}
\pmatrix{0&m_D\cr m_D^T&M_R}
\end{equation}
where $m_D$ is an $n\times n$ matrix of Dirac mass terms,
 and $M_R$ is an $n\times n$ matrix of Majorana mass terms.
We can now see in what context the parameter $\alpha$ defined by
eq.~(\ref{eq:pofmu})
takes its limiting values. For simplicity, consider the situation $n=1$.
In order to find the physical states the matrix (\ref{eq:massmatrix}) must be
diagonalised.
In the limit $M_R>>m_d$ we have a purely-active Majorana field with
 mass eigenvalue  $\sim m_d^2/M_R$; and a purely-sterile Majorana field with
mass eigenvalue $M_R$. In the naive  see-saw model $M_R$ is set at a
large enough value so as to decouple the sterile fields from the theory.
In the opposite scenario where $M_R<<m_D$, the mass eigenvalues
become $m_D\pm M_R$. Since in this case the eigenvalues are almost degenerate,
the
neutrino field is described by four degrees of freedom and  for all practical
purposes behaves as  a Dirac field.

For pure Dirac (Majorana)  neutrino fields and spin-zero bosons,
 $\alpha=-1 (0)$.
This can be understood heuristically as follows:
for a heavy right-handed fermion to decay into
a light left-handed fermion plus a boson,
either the fermion spin or the fermion momentum must flip.  If the boson
is spin-zero, then the fermion spin cannot be flipped because of
angular momentum conservation, hence the fermion momentum must flip.
Thus the boson is preferentially forwards produced, and cannot be emitted
in the backward direction, $\alpha=-1$.  If, however, the light fermion is
a Majorana fermion, then it is it's own antiparticle hence one must
include the process where the final state fermion is a right-handed
anti-neutrino.  The decay to this state involves no helicity flip, and
hence the angular dependence is exactly opposite, $\alpha = +1$.
Taken together, the decay is isotropic  (Li and Wilczek 1982; Shrock 1982).
We then have the limiting cases
\begin{equation}
\alpha\rightarrow 0 \>{\rm for} \>  M_R>>m_D
\end{equation}
\begin{equation}
\alpha\rightarrow -1\> {\rm for} \  M_R<<m_D\>.
\end{equation}

Let us now consider the coupling of the Majorana neutrino fields to the
majoron $\chi$.
For clarity we will adopt $n=2$: generalisation to $n=3$ is
straightforward.
For $n=2$, our Lagrangian has the explicit form
\begin{equation}	\label{eq:massmatrix2gen}
\overline\nu_{w}
\pmatrix{0&0&g_{ee}v&g_{e\mu}v\cr
0&0&g_{\mu e}v&g_{\mu\mu}v\cr
g_{ee}v&g_{\mu e}v&h_{ee}u&h_{e\mu}u\cr
g_{e\mu}v&g_{\mu\mu}v&h_{e\mu}u&h_{\mu\mu}u\cr}\nu_{w}
+
\overline\nu_{w}
\pmatrix{0&0&0&0\cr
0&0&0&0\cr
0&0&h_{ee}u&h_{e\mu}u\cr
0&0&h_{e\mu}u&h_{\mu\mu}u\cr}{i\over 2}\chi\nu_{w}
\>
\end{equation}
where we label our weak  eigenstates as
\begin{equation}
\nu_w=\pmatrix{\nu_{e(L)}\cr \nu_{\mu (L)}\cr \nu_{e(R)}\cr \nu_{\mu (R)}\cr
}\> .
\end{equation}
To determine the couplings of the neutrino fields to the $\chi$ field,
we must first diagonalise the first $4\times 4$ matrix of
eqn.(\ref{eq:massmatrix2gen})  (the mass matrix),
in order to establish the physical states. In the simplified scenario
where there are no mixed terms ($g_{\alpha\beta}=h_{\alpha\beta}=0$ for all
 $\alpha\neq \beta$)
each neutrino generation will be described by two
physical fields
\begin{eqnarray}
P_1=\sin\theta \ \nu_{L} +\cos\theta  \ \nu_{R}\cr
P_2=-\cos\theta \  \nu_{L}+ \sin\theta \ \nu_{R}\>,\cr
\end{eqnarray}
where the rotation angle $\theta$ is given by
\begin{equation}
\tan 2\theta={2g_{\alpha\alpha}v\over h_{\alpha\alpha}u} \ \ \ ; \ \sin
2\theta={2g_{\alpha\alpha}v\over
[h_{\alpha\alpha}^2u^2+4g_{\alpha\alpha}^2v^2]^{1/2}}\>.
\end{equation}
%The masses of the states are
%\begin{eqnarray}
%M_1 = {{\sqrt {h_{\alpha\alpha}^2 u^2 + 4g_{\alpha\alpha}^2 v^2}}
%     + h_{\alpha\alpha} u \over 2} \>;
%M_2 = {{\sqrt {h_{\alpha\alpha}^2 u^2 + 4g_{\alpha\alpha}^2 v^2}}
%     - h_{\alpha\alpha} u \over 2}
%\end{eqnarray}
The coupling of these physical fields to the majoron field is then given by
\begin{equation}
[\overline P_1,\> \overline P_2]\pmatrix{\cos^2\theta&\sin 2\theta\cr
\sin 2\theta& \sin^2\theta\cr} {ih_{\alpha\alpha}u\over 2}\chi \pmatrix{P_1 \cr
P_2\cr}
\>.
\end{equation}

It is clear that this simplest scheme allows for possible decay modes of the
form $P_2\rightarrow P_1+\chi$ with a decay rate set by the coupling constants
and the vacuum expectation values,  $<H>$ and $<S>$. In the limit that $\theta$
is small, this can  be loosely  characterised as the decay of sterile  neutrino
field into
an active neutrino field with the emission of a majoron.
With the introduction of mixed terms ($g_{\alpha\beta}\ {\rm and} \
h_{\alpha\beta}\ne0$), the generalisation to $n=3$, and the freedom to vary
mass terms; a rich variety of decay schemes becomes possible.
This is especially so if more complicated singlet majoron models
are introduced (eg. Berezhiani, Smirnov and Valle 1992; Burgess and Cline
1993),
 or if more than one new singlet field $S$
is utilised.
However,  it can be seen that a decay of the form
\begin{equation}
\nu_{\tau(L)}\rightarrow\nu_{\mu(R)}+\chi
\end{equation}
is possible if $m_{\nu_\mu(R)} < m_{\nu_\tau(L)}$.
Since this form of decay is possible in this simple
extension of the neutrino sector, and does not involve a new type of fermion,
it may be a viable framework in which to try to build a complete model
incorporating MDM production by neutrino lasing.

We also note that heavy sterile fields in the singlet-majoron model
can  have  decays of the form
$\nu_{\tau(R)}\rightarrow\nu_{e(R)}+\chi$. If, for example by imposition of
discrete symmetries (eg. Carlson and Hall 1989), these decays are slower
 than the decay
of the $\nu_{e(R)}$ field, a scenario for neutrino lasing arises.
Here the $\nu_{e(R)}$  number density will  come into chemical equilibrium and
 be reduced relative to the number
density of $\nu_{\tau(R)}$'s, and  phase space will be opened up for the
heavier $\nu_{\tau(R)}$ to undergo lasing (likewise for $\nu_{\mu(R)}$).

Detailed
investigation of  decay schemes such as those outlined above would be of value
in examining the feasibility of neutrino lasing. In this spirit, models of
other
decay schemes such as those embedded in
  more complicated
singlet-majoron models would also be useful.
We remind would-be model builders, however, that
 the possibility of   relevant decays  is a necessary but not sufficient
condition to be  satisfied --
one must still satisfy  other conditions:
1) Decoupling of the light fermion in such a way as to produce a population
inversion
 (with no reactions other than the decay remaining active);
2) $T_d$ and $m_H$ in the lasing region of parameter space;
3) a boson mass of $\sim 30$eV, while avoiding cosmological limits
such as from early universe production  of domain walls;
4) accommodation of all experimental bounds on neutrino and weak interaction
physics in general.

\section{Variations on a Theme} \label{sec:variations}

\subsection{Decay Processes Involving Known Neutrinos}

\subsubsection{$\nu_\tau\rightarrow \nu_\mu +B$} \label{sec:scenario2}

We have shown how a significant production of low-momentum bosons may take
place in
the early universe. A vital key to this production was the fact that F was not
one of the known neutrinos. Let us briefly mention here the situation if F were
in fact identified
with one of the known neutrinos.
To be specific let us consider the decay process $\nu_\tau\rightarrow \nu_\mu
+B$,
where we have implicitly assumed $m(\nu_\tau) > m(\nu_\mu)$.
The assumption here is that quark confinement creates the light and heavy
neutrinos at equal rates, and that the bosons decoupled prior to the QCD phase
transition.
 Therefore
 the initial conditions are
$\mu_i = 0$, $T_F^i = T_\tau^i = {\left({{g_b}/{g_a}}\right)}^{1/3} T_B^i$,
where $g_i=\sum (g_b +{7 \over 8} g_f)$ is the statistical weight
 of relativistic particles
 in bosons, $g_b$, and  in fermions, $g_f$ (subscripts $a$ and $b$
 refer to epochs
 after and before the annihilation epoch).
 Such decoupling of the bosons is required otherwise they would
 take up their
full statistical weight;  which
in terms of additional neutrino degrees
of freedom can be written
 $\Delta N_\nu=4/7 (8/7)$ for B$=\bar{\rm B}$ (B$\ne\bar{\rm B}$).
Such a large boson abundance
 would violate the  primordial nucleosynthesis bound
on relativistic degrees of freedom:
$\Delta N_\nu\le 0.3$~.

There is now no population inversion, lasing does not take place, and
it follows
from the
Boltzmann equations (\ref{eq:boltzmann}) that very little will happen until
$T_d$.
In this case
 it is  physically reasonable to have the decay processes
commencing at some arbitrary time, determined by the value of the
free decay time. If the decay occurs at
 $T>2.3$ MeV, then since the neutrinos are  in
chemical  equilibrium, the bosons will take up their
full statistical weight,  again violating the  primordial nucleosynthesis
bound.

Consider instead the situation where the decay processes commence
at $T< 2.3$ MeV and
statistical equilibrium is no longer established.
At $T_d$, while the decays cannot establish
statistical equilibrium,
the three components of the system will  come into chemical equilibrium,
assuming all  particles are   relativistic, just as calculated by Madsen.
Now it is true the coupling to very low momentum bosons ($\pb/T \ll (m^2/T^2)$)
is exponentially suppressed, so the decays cannot truly establish
equilibrium, but
this is not a serious consideration since in this case one does not find
a Bose condensate anyway.

Moreover, only about 20\% of the $\nut$'s have decayed.
The $\nut$'s are still relativistic.
The equilibrium state remains valid until
the $\nut$'s go non-relativistic when
the remaining $\nut$'s decay.
The net result is a non-thermal spectrum, but with no particularly large
low-momentum component.

\subsubsection{QCD Lasing: $\nu_\mu\rightarrow \nu_\tau +$ B}
			\label{sec:qcdlasing}

\def \Gammaweak {{\Gamma_{\rm W}}}
\def \Tweak {{T_{\rm W}}}
Another alternative to having the F be an unknown degree of
freedom, is to identify H with $\nu_\mu$, and F with $\nu_\tau$ but
with $m(\nu_\mu) > m(\nu_\tau)$, so that $\nu_\mu \to \nu_\tau + $ B is
possible.
As discussed above, during the quark confinement phase transition, which
takes place when
$T_{QCD} \sim 1-300\MeV$,
quark degrees of freedom are being eliminated
resulting in a net production of
those other degrees of freedom
whose number changing interactions are not yet frozen out,
in particular the three known left-handed neutrinos.
We can divide the processes which produce these $\nu$'s into two types:
charged current processes, in which the $\nu$'s are produced
together with (the antiparticle of ) their associated charged lepton
($\nu_e$ with ${\bar e}$, $\nu_\mu$ with ${\bar \mu}$, \etc);
and neutral current processes in which they are produced
in association with their own anti-particle
($\nu_e$ with ${\bar \nu_e}$, \etc).
Since the mass of the $\tau$-lepton is $1.78\GeV$, which is $\gg T_{QCD}$,
whereas the mass of the muon is only $105\MeV \lsim T_{QCD}$,
charged current production of $\nu_\tau$'s is strongly suppressed
compared with charged current production of $\nu_\mu$.  This produces a small
excess of $\nu_\mu$ compared to $\nu_\tau$,
\ie~ the reheating of the $\nu_\tau$ lags behind that of the $\nu_\mu$..

   In the absence of the decay $\nu_\mu \to \nu_\tau + $ B,
weak interactions (\eg~ $\nu_\mu {\bar \nu_\mu} \to \nu_\tau {\bar \nu_\tau} $)
act to erase the population inversion and maintain equilibrium.  This is
quite efficient since $\Gammaweak \sim 10^{4.5} H $ at $T=100\MeV$, where $H$
is the Hubble parameter.
Nevertheless, $f_{\nu_\tau} = f_{\nu_\mu}( 1 - {\cal O}(H / \Gammaweak))$,
so if $m(\nu_\mu) > m(\nu_\tau)$,
then this $(f_{\nu_\tau} - f_{\nu_\mu})$ constitutes a population inversion.
If $\Gammalase \gsim \Gammaweak$ then the lasing process will compete with
(or dominate) the weak interactions in erasing the population inversion,
resulting in the production of cold bosons.
In fact if $\Gammalase \gg H$ then the lasing process will exponentiate
until the net rate of lasing events becomes roughly equal to the rate
at which excess $\nu_\mu$'s are injected.
(There is also an excess of $\nu_e$ over $\nu_\mu$, because
$m_e<<m_\mu\simeq\TQCD$ but if B is to be a candidate dark matter particle
then $m_B \gsim 10$eV, whereas $m(\nu_e) \lsim 7eV$, so there is
no possibility of $\nu_e$ decays producing B's.)

Unfortunately, lasing during the quark confinement phase transition tends
to produce a small fraction of cold bosons, because charged
current production contributes at most $50\%$ of the $\nu_\mu$'s.

\subsection{Non-Standard Nucleosynthesis}

Finally, we should mention at this point,
the possibility of directly affecting the
primordial nucleosynthesis  by the $\nu_\tau$ decay
in the circumstance where F$=\nu_e$, as pointed out by Enqvist {\it et al.}
(1992) and Madsen (1992). The point here is that if the decay happens to
occur in the narrow window between chemical decoupling at $2.3$ MeV and
thermal decoupling at $0.7$ MeV, the increase in the $\nu_e$ and $\bar\nu_e$
chemical potentials can directly influence the weak interactions
maintaining the neutron and proton equilibrium. The net result of this would be
a significant {\it reduction} in the $^4$He yield. Although
this requires fine tuning of the decay process, it does alleviate  the
concerns with regard to $\Delta N_\nu$.
Also we note that
SBBN makes a clear prediction  for the $^4$He mass fraction:
 $Y_p>0.237$ (Smith {\it et al.} 1993);
 future  observational
determinations of  an upper limit to $Y_p$ below  this level
would render modifications to the SBBN theory mandatory. In this case,
a plethora of new nucleosynthesis models stand in the wings
(for a review of non-standard models see Malaney and Mathews 1993),
and
 limits on $\Delta N_\nu$ would be subject to change.
This last point would be of most relevance
to scenarios where F$\ne\nu_e, \nu_\mu$.

\section{CONCLUSIONS}

We have investigated the possibility of producing
non-thermal particle distributions in the early universe
from decay processes of the form H $\rightarrow$ F $+$ B,
where H is probably one of the known neutrinos.
In the case where F is also one of the known neutrinos,
the departure from thermal equilibrium distributions is not as dramatic
as the case where F is some as yet undiscovered particle
which was decoupled from the primeval plasma at $T_{QCD}$.
In this latter case,
the possibility arises of forming a  quasi-Bose-condensate --
a boson momentum distribution where a significant fraction of
the bosons are in low momentum states.
We have detailed how the efficiency of producing cold bosons is
influenced by the spin of the participating particles,
and how the final distributions depend on position in the
$(\mh,\mb,T_d)$ parameter space. We have attempted to identify a
particle physics framework within which
the required decay scheme could be constructed.

The implications of the $\nu$-lasing scenario for large-scale
structure depend on two parameters: the ratio of cold to hot
particles and the effective temperature of the hot particles.
We have emphasised here how the former is sensitive to details of the
particle physics model such as the number of decay
channels.

A thorough assessment of the viability of these models will require
numerical solution of the Boltzmann equations as input to numerical
calculation of the transfer function which, in turn, will provide the
input for non-linear evolution via
N-body simulation.  This is beyond the scope of this paper.

Future observations are likely to discriminate between the different
dark-matter phenomenologies. If such observations definitely determine that
modifications
to the standard CDM model are required, and that MDM is the preferred choice,
then a single
mechanism for producing multi-component momentum distributions
will clearly be of import.

\acknowledgements

We thank M.\ Butler,  S. Doddleson, B.\ Holdom, J.\ Madsen, M.\ Savage and L.\
Widrow
for helpful discussions.
\vfill
\eject
\centerline{FIGURE CAPTIONS}
\medskip
\noindent
Figure 1. Kinematic constraints on $\Eb$, $\Eh$ (logarithmic scale).

\medskip
\noindent
Figure 2a.
 $\alpha\neq -1$ (low-momentum boson emission unsuppressed).
The parameter space is divided into five regimes:
(i) $\mh \leq \mb$, no decays;
(ii) $T_d \lsim (\mb\mh)^{1/2}$ decays, no lasing;
(iii) $(\mb\mh)^{1/2}\lsim T_d\lsim\Estar$ and $T_{lase}\lsim 2$MeV,
lasing, no post-lasing thermalisation;
(iv) $T_d\gsim\Estar$, $T_{lase}\lsim 2$MeV, lasing, post-lasing
thermalisation;
(v) $T_{lase}\gsim 2$MeV, lasing raises helium-4 abundance unacceptably.

\medskip
\noindent
Figure 2b. Parameter space for lasing for $\alpha\simeq -1$ (low-momentum boson
emission suppressed).
The parameter space is divided into five regimes:
(i) $\mh \leq \mb$, no decays;
(ii) $T_d \lsim \mh$ decays, no lasing;
(iii) $\mh\lsim T_d\lsim\Estar$ and $T_{lase}\lsim 2$MeV,
lasing, no post-lasing thermalisation;
(iv) $T_d\gsim\Estar$, $T_{lase}\lsim 2$MeV, lasing, post-lasing
thermalisation;
(v) $T_{lase}\gsim 2$MeV, lasing raises helium-4 abundance unacceptably.

\vfill
\eject


\begin{references}
\reference

Berezhiani,  Z. G., Smirnov, A. Yu. and Valle , J. W. F., 1992, {\it Phys.
Lett. B}, {\bf 291}, 99.
\reference
Burgess, C. P. and Cline , J., 1993, {\it Phys. Lett. B}, {\bf 298}, 141.
\reference
Carlson, E. D. and Hall, L., 1989, {\it Phys Rev. D}, {\bf 40}, 3187.
\reference
Chikashige, Y., Mohapatra, R. N., and Peccei, R. D., 1981, {\it Phys. Lett. B},
{\bf 98}, 265.
\reference
Davis, M., Summers, F. J.  and Schlegel, D., 1992, {\it Nature}, {\bf 359},
393.
\reference
Enqvist, K, Kainulainen, K. and Thompson, M., 1992, {\it Phys. Rev. Lett.},
{\bf 68}, 744.
\reference
Feldman, H., Kaiser, N., and Peacock, J. 1993.  {\it Ap. J.}, in press.
\reference
Feynman, R.P. Leighton, R.B. and Sands, M., 1965.
{\it Lectures on Physics}, Vol 3, Addison-Wesley.
\reference
Fisher, K.B., Davis, M., Strauss, M.A., Yahil, A. and Huchra,
J.P., 1993 {\it Ap. J.}, {\bf 402} 42.
\reference
Kaiser, N., Malaney, R. A. and Starkman, G. D., 1993. {\it Phys. Rev. Lett.},
{\bf 71}, 1128.
\reference
Kawasaki, N., Steigman, G. and Kang, H. S., 1993, {\it Nuc. Phys. B}, {\bf
403}, 671.
\reference
Klypin, A., Holtzman, J., Primack, J. and Reg\"os, E., 1993, preprint.
\reference
Kolb, E. W. and Turner, M. S.,  1990, {\it The Early Universe},
 (Addison-Wesley; Redwood City).
\reference
Li, L.F. and Wilczek, F., 1982, {\it Phys. Rev. D}, {\bf 25}, 143.
\reference
Maddox,  S. J., Efstathiou, G. P.,  Sutherland , W. and Loveday, J., 1990,
{\it Mon. Not. R. Astr. Soc.,} {\bf 242}, 43.
\reference
Madsen, J., {\it Phys. Rev. Lett.}, 1992, {\bf 69}, 571.
\reference
Malaney, R. A. and Mathews , G. J., 1993, {\it Physics Reports}, {\bf 229},
Number 4, 145.
\reference
Ostriker,  J. P, 1993, {\it Ann. Rev. Astron. Astrophys.}, {\bf 31}, 689.
\reference
Pagel,  P. E. J., Simonson, E. A., Terlevich, R. J. and Edmunds, M. G., 1992,
{\it M. N. R. A. S.}, {\bf 255}, 325.
\reference
Schaefer, R., Shafi, Q., and Stecker, F., 1989, {\it Ap. J.}, {\bf 347}, 575.
\reference
Schaefer, R., and Shafi, Q., 1992. {\it Nature}, {\bf 359}, 199.
\reference
Shafi, Q. and Stecker, F., 1984, {\it Phys. Rev. Lett.}, {\bf 53}, 1292.
\reference
Shrock, R.E. , 1982, {\it Nucl. Phys. B}, {\bf 206}, 359.
\reference
Smith, M.,  Kawano, L. and Malaney, R. A., 1993,
{\it Ap. J. Supp.}, {\bf 85}, 219.
\reference
Smoot, G.  {\it et al.}, 1992, {\it Ap. J. Lett.}, {\bf 396}, L1.
\reference
Taylor, A. N.  and Rowan-Robinson, M.,  1992,
 {\it Nature}, {\bf 359}, 396.
\reference
Vogeley, M.S., Park, C., Geller, M. \& Huchra, J.P., 1992, {\it Ap. J. Lett.,}
{\bf 391} L5.
\reference
Wagoner,  R. V., 1979, in {\it Physical Cosmology- Les Houches 1979}, eds.
Balian, R., Audouze, J. and Schramm, D. N., p395 (North Holland; Amsterdam).
\reference
Walker, T. P., Steigman, G., Schramm, D. N., Olive, K. A. and Kang, H. S.,
1991,
{\it Ap. J.}, {\bf 376}, 393.
\reference
Wright, E. L.  {\it et al.}, 1992, {\it Ap. J. Lett.,}
 {\bf 396}, L13.




\end{references}
\end{document}